\documentclass[RNAAS]{aastex62}

\begin{document}

\title{Photometry and preliminary modelling of Type IIb Supernova 2017gpn}

%% Note that the corresponding author command and emails has to come
%% before everything else. Also place all the emails in the \email
%% command instead of using multiple \email calls.
\correspondingauthor{Maria V. Pruzhinskaya}
\email{pruzhinskaya@gmail.com, buzm-ea@yandex.ru, moskvitin.alexander@gmail.com, sergei.blinnikov@itep.ru}

\author{Maria V. Pruzhinskaya}
\affiliation{Lomonosov Moscow State University, Sternberg Astronomical Institute, Universitetsky pr.~13, Moscow, 119234, Russia}
\affiliation{Space Research Institute, 84/32 Profsoyuznaya Street, Moscow, 117997, Russia}

\author{Elena A. Balakina}
\affiliation{Lomonosov Moscow State University, Sternberg Astronomical Institute, Universitetsky pr.~13, Moscow, 119234, Russia}
\affiliation{Lomonosov Moscow State University, Faculty of Physics, Leninskie Gory, 1-2, Moscow, 119991, Russia}

\author{Alexander S. Moskvitin}
\affiliation{Special Astrophysical Observatory RAS, Nizhnij Arhyz, 369167, Russia}

\author{Sergey I. Blinnikov}
\affiliation{Lomonosov Moscow State University, Sternberg Astronomical Institute, Universitetsky pr.~13, Moscow, 119234, Russia}
\affiliation{Space Research Institute, 84/32 Profsoyuznaya Street, Moscow, 117997, Russia}
\affiliation{NRC "Kurchatov institute" - ITEP, B.Cheremushkinskaya 25, 117218 Moscow, Russia}
\affiliation{Kavli IPMU, University of Tokyo, Kashiwa, 277-8583, Japan}

\keywords{supernovae: general -- supernovae: individual: SN 2017gpn -- stars: evolution}

\section{Introduction} 

During follow-up inspection of the error box of the LIGO/Virgo alert G299232 on 2017 August 27.017, MASTER Global Robotic Net~\citep{2010AdAst2010E..30L} discovered an optical transient named MASTER OT~J033744.97+723159.0~\citep{GCN21780,roberts2018}. 

On 2017 August 29, the spectrum of MASTER OT J033744.97+723159.0 was obtained with the Xinglong 2.16-m telescope of National Astronomical Observatory of China~\citep{2017ATel10681....1R}. The object was classified as Type IIb Supernova (SN) by cross-correlating with a library of spectra (SNID;~\citealt{2007ApJ...666.1024B}).

On 2017 September 6, M.~Caimmi reported the discovery of a supernova with 0.24-m telescope from Valdicerro Observatory~\citep{2017TNSTR.973....1C}. The supernova received the IAU designation AT 2017gpn and was identified as MASTER OT J033744.97+723159.0.

\section{Observations and data reduction}
We performed 20 epochs of observations ($B$ and $R$ filters) with CCD-photometer on the Zeiss-1000 telescope of the Special Astrophysical Observatory of the Russian Academy of Sciences.
The aperture photometry was performed using standard procedures of ESO-MIDAS software package. It includes standard image processing such as bias subtraction and flat field correction, removing the traces of cosmic particles, and stacking of individual frames into the summary image. 
SN 2017gpn is located in $\sim$0.039 degrees ($>20$ kpc) from the center of the potential host galaxy NGC1343, so the galaxy's contamination is negligible. The line-of-sight reddening is adopted to be $E(B - V) = 0.017$ mag~\citep{2011ApJ...737..103S}. Since no Landolt or any other standards were available for this region, we use the Pan-STARRS magnitudes for comparison stars. The magnitudes of comparison stars were re-calculated from $g,r,i$ to $B,R$ with use of Lupton  transformation equations\footnote{http://www.sdss3.org/dr8/algorithms/sdssUBVRITransform.php}. 

For better modelling we combined our photometric data with publicly available $B$ and $R$ light curves (LCs) of SN~2017gpn obtained with PIRATE robotic telescope~\citep{roberts2018}.

\section{Modelling}
The numerical light curve modelling is performed with one-dimensional multifrequency radiation hydrodynamical code \textsc{STELLA}. The full description of the code can be found in~\cite{Blinnikov1998,Blinnikov2006}; a public version of \textsc{STELLA} is also included with the \textsc{MESA} distribution~\citep{Paxton2018}. 

Our best-fit numerical model is shown by solid line in the Fig.~\ref{fig:model}. The parameters of the model are: pre-SN mass $M = 3.5~\rm M_\odot$, pre-SN radius $R = 50~\rm R_\odot$, mass of hydrogen envelope $M_{env} = 0.06~\rm M_\odot$. The explosion energy is $E = 1.2\times10^{51}$ erg. The 0.11~$\rm M_\odot$ of $^{56}$Ni is totally mixed through the ejecta. The compact remnant is $1.41~\rm M_\odot$ neutron star.

The theoretical photospheric velocities are in a good agreement with the observed one ($\sim$14800 km~s$^{-1}$ from SiII~635.5nm absorption line, \citealt{2017ATel10681....1R}). 

\section{Discussion and conclusions}
\begin{figure}[h!]
\begin{center}
\includegraphics[scale=0.75,angle=0]{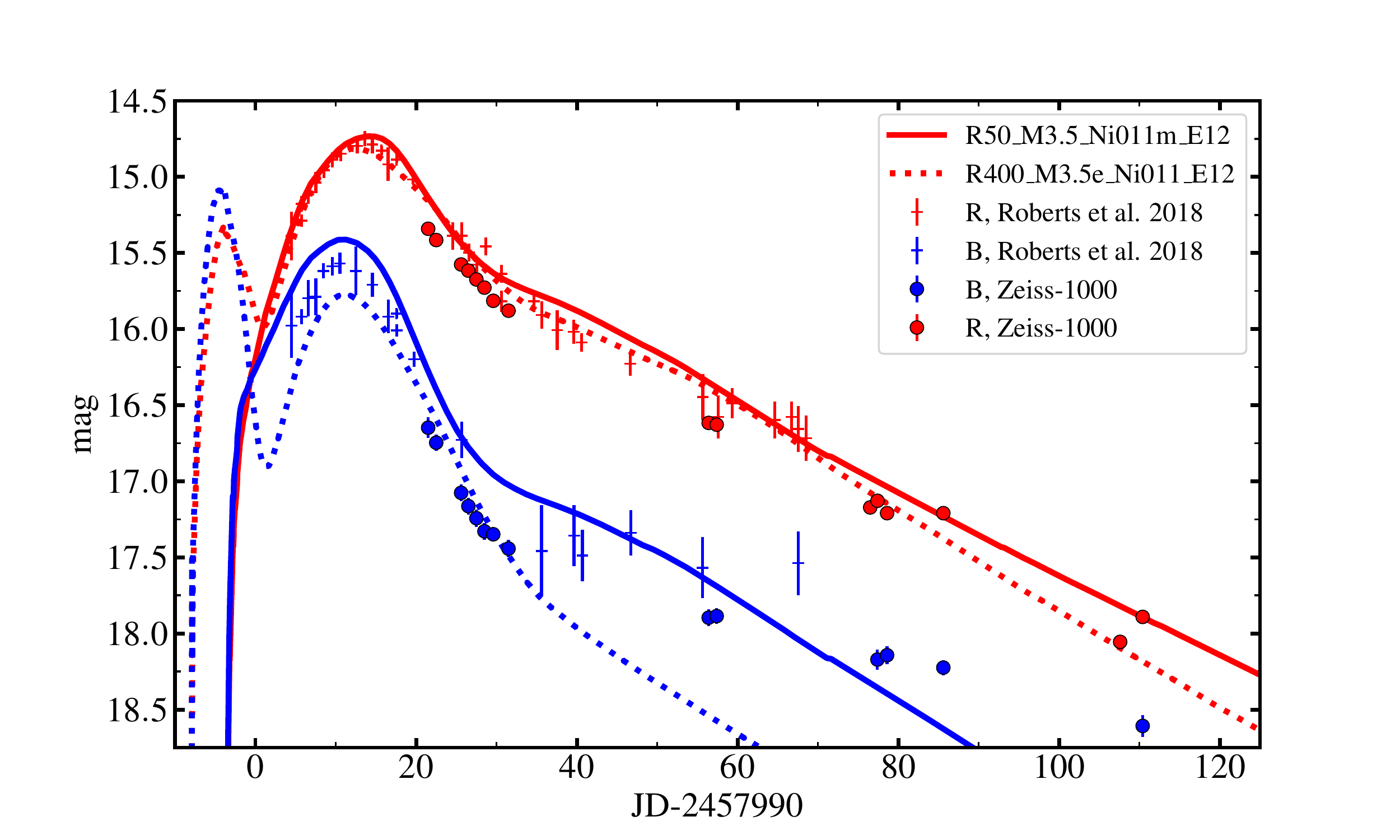}
\caption{The result of modelling the observed LCs of SN~2017gpn (points are our data and crosses are the data taken from~\citealt{roberts2018}). Our best-fit model is shown by solid lines ($M = 3.5~\rm M_\odot$, $R = 50~\rm R_\odot$, $E = 1.2\times10^{51}$ erg,  $M_{^{56}Ni} = 0.11~\rm M_\odot$, mixed). For comparison the model with  $R = 400~\rm R_\odot$ ($M_{^{56}Ni} = 0.11~\rm M_\odot$, no mixing) is presented.\label{fig:model}}
\end{center}
\end{figure}

The parameters we found are consistent with the results of hydrodynamical modelling of other typical Type IIb supernovae. However, in different hydrodynamical models of SN~IIb there is some variance in radius of pre-SN star (from 30-50~R$_\odot$ to 700~$R_\odot$, e.g.,~\citealt{1994ApJ...429..300W,Blinnikov1998,2015ApJ...811..147F}). To check if it is possible to reproduce the observed LCs of SN~2017gpn with a model of higher radius, we changed the radius in our best-fit model to $R=400~\rm R_\odot$ and variated the degree of $^{56}$Ni mixing. Concentrating all the $^{56}$Ni in the central part of ejecta, we were able to nearly reproduce the observed LCs~(Fig.~\ref{fig:model}). This stresses the importance of $^{56}$Ni mixing in such kind of studies.

By adopting the date of explosion from the models (Aug 20 for our best-fit model and Aug 16 for the model with $R=400~\rm R_\odot$), we can conclude that SN~2017gpn is unlikely to be connected to the LIGO/Virgo G299232 alert.

\acknowledgments
M.V.P. and S.I.B. acknowledge support from RSF grant 18-12-00522 for modelling. A.S.M. is grateful to O.I.~Spiridonova and the Zeiss-1000 staff for the help in observations.

\bibliographystyle{aasjournal}
\bibliography{biblio}

\end{document}